\documentclass{PoS}

\usepackage{amsmath,amssymb}
\usepackage{caption}
\DeclareMathAlphabet{\mathcal}{OMS}{cmsy}{m}{n}
\title{Tensor network study \\ of two dimensional lattice $\phi^{4}$ theory}

\ShortTitle{TN study of 2D $\phi^{4}$ theory}

\author{\speaker{Ryo Sakai}\thanks{This work was supported by the Ministry of Education, Culture, Sports, Science and Technology (MEXT) as ``Exploratory Challenge on Post-K computer'' (Frontiers of Basic Science: Challenging the Limits) and JSPS KAKENHI Grant Numbers JP16K05328, JP17K05411, JP18J10663.},\quad Shinji Takeda$^{\dagger}$ \\
  Institute for Theoretical Physics, Kanazawa University, Kanazawa 920-1192, Japan \\
  E-mail: \email{sakai@hep.s.kanazawa-u.ac.jp},\quad \hspace{-0.1em}\email{takeda@hep.s.kanazawa-u.ac.jp}}

\author{Daisuke Kadoh$^{\dagger}$ \\
  Department of Physics, Faculty of Science, Chulalongkorn University, Bangkok 10330, Thailand, \\
  Research and Educational Center for Natural Sciences, Keio University, Yokohama 223-8521, Japan \\
  E-mail: \email{kadoh@keio.jp}}

\author{Yoshinobu Kuramashi$^{\dagger}$,\quad Yusuke Yoshimura \\
  Center for Computational Sciences, University of Tsukuba, Tsukuba 305-8577, Japan \\
  E-mail: \email{kuramasi@het.ph.tsukuba.ac.jp},\quad \email{yoshimur@ccs.tsukuba.ac.jp}}

\author{Yoshifumi Nakamura\\
  RIKEN Center for Computational Science, Kobe 650-0047, Japan\\
  E-mail: \email{nakamura@riken.jp}}

\abstract{
  The tensor renormalization group attracts great attention as a new numerical method that is free of the sign problem.
  In addition to this striking feature, it also has an attractive aspect as a coarse-graining of space-time;
  the computational cost scales logarithmically with the space-time volume.
  This fact allows us to aggressively approach the thermodynamic limit.
  While taking this advantage, we study the critical coupling of the two dimensional $\phi^{4}$ theory on large and fine lattices.
  We present the numerical results along with the extrapolation procedure to the continuum limit and compare them with the previous ones by Monte Carlo simulations.
}

\FullConference{The 36th Annual International Symposium on Lattice Field Theory - LATTICE2018\\
  22-28 July, 2018\\
  Michigan State University, East Lansing, Michigan, USA.}

\begin{document}

\section{Introduction}
\label{sec:Introduction}

The $\phi^{4}$ theory is a scalar field theory which has only the self quartic interaction term,
and it is regarded as the simplest interacting model.
Despite the simplicity, the theory plays important roles in particle physics,
and there are non trivial features surrounding the model.
One of its important features is the spontaneous $Z_{2}$ symmetry breaking.
The symmetry breaking is closely related to the phase transition,
and our aim in this report is to precisely analyze a critical phenomena of this model.

This model displays the different behaviors depending on its dimension.
In five dimensions or higher, it goes to the free theory in the continuum limit,
and this property, namely the triviality of the $\phi^{4}$ theory, is believed to hold also in four dimensions.
By contrast, the two and three dimensional $\phi^{4}$ theory in the continuum limit has finite coupling constant,
and many numerical calculations have been done to determine the non-trivial coupling constant.
In two dimensions many numerical approaches have been reported,
and recently precise computations have been developing using celebrated algorithms such as cluster algorithm and the worm algorithm~\cite{Schaich:2009jk,Wozar:2011gu,Bosetti:2015lsa,Bronzin:2018tqz}.
From this background, computation of the critical coupling can be regarded as a benchmark test for numerical algorithms.

In this report we compute the dimensionless coupling using the tensor renormalization group (TRG).
The TRG is a coarse-graining procedure for space-time,
and this feature is useful for taking the thermodynamic limit.
In short, the computational time of the TRG is in proportion to the logarithm of space-time volume.
This is a big advantage compared to Monte Carlo simulations, whose computational time is proportional to the space-time volume.
Taking this advantage we explore the continuum limit of the theory and try to precisely determine the value of the critical coupling.

This report is organized as follows.
We first present the definition of the model and a tensor network representation for the target quantity in sec.~\ref{sec:2Dphi4_TN}.
In sec.~\ref{sec:numericalresults}, we show the numerical results along with the extrapolation procedure to the continuum limit.
Section~\ref{sec:Summary} is devoted to summary and discussion.

\section{Two dimensional lattice $\phi^{4}$ theory and its tensor network formulation}
\label{sec:2Dphi4_TN}

The action of the $\phi^{4}$ theory in two dimensions is defined as
\begin{align}
  \label{eq:phi4_action}
  S_{\mathrm{cont.}}
  = \int \mathrm{d}^{2}x
  \left\{
  \frac{1}{2} \left( \partial_{\nu} \phi \right)^{2}
  + \frac{\mu_{0}^{2}}{2} \phi^{2}
  + \frac{\lambda}{4} \phi^{4}
  \right\}
\end{align}
with the bare mass $\mu_{0}$ and the bare coupling $\lambda$.
$\phi$ denotes one-component real scalar field: $\phi \in \mathbb{R}$.
This model is obviously invariant under the $Z_{2}$ transformation ($\phi \rightarrow -\phi$);
however, this symmetry may be dynamically broken in the context of quantum theory.
Then the expectation value of the scalar field is regarded as an order parameter;
$\left<\phi\right> = 0$ corresponds to the symmetric phase, and if it takes a nonzero value, the system is in the symmetry broken phase.

As a non-perturbative method, the lattice formulation is useful to study such dynamics.
Thus, from here let us work on the square lattice.
Since $\mu_{0}^{2}$ and $\lambda$ have positive mass dimension,
the dimensionless form of the parameters are introduced as
\begin{align}
  \label{eq:dimensionlessparameters}
  \hat{\mu}_{0}^{2} = a^{2} \mu_{0}^{2}, && \hat{\lambda} = a^{2} \lambda
\end{align}
with the lattice spacing $a$.
In the following we suppress the hat for simplicity;
otherwise one can think that the lattice units $a=1$ is assumed.
After the preparation, the two dimensional lattice action is written as
\begin{align}
  \label{eq:S_externalfield}
  S_{h}
  = \sum_{n} \left\{
  \frac{1}{2} \sum_{\nu=1}^{2} \left( \phi_{n+\hat{\nu}} - \phi_{n} \right)^{2}
  + \frac{\mu_{0}^{2}}{2} \phi_{n}^{2}
  + \frac{\lambda}{4} \phi_{n}^{4}
  - h \phi_{n}
  \right\},
\end{align}
where $n$ denotes the coordinate in $L \times L$ space-time lattice and $\hat{\nu}$ denotes the unit vector along the $\nu$-axis.
Note that in eq.\eqref{eq:S_externalfield}, the external field $h$ is introduced
though it is absent in the continuum version in eq.~\eqref{eq:phi4_action}.
As will seen, the external field plays some roles to
extract a non-zero value of the expectation value of the scalar field in the thermodynamic limit.

Our final goal in this report is to evaluate the dimensionless critical coupling of the $\phi^{4}$ theory.
For that purpose, we need to calculate physical quantities.
Here we introduce the tensor network representation of the partition function
\begin{align}
  \label{eq:phi4_partitionfunction}
  &Z
    = \left(\prod_{n} \int \mathrm{d}\phi_{n} \right) e^{-S_{h}},
\end{align}
where $n$ in the product runs all lattice sites, following our previous work~\cite{Kadoh:2018hqq}.

To get straight to the point, we discretize the scalar field using $K$ nodes to construct the tensor network formulation~\footnote{
  The detail procedure is given in ref.~\cite{Kadoh:2018hqq},
  and the systematic errors from discretization of scalar fields are also discussed there.
},
and the partition function is expressed as a product of tensors
\begin{align}
  \label{eq:Z_tensornetwork_approx}
  Z\left( K \right)
  = \left( \prod_{n} \sum_{x_{n}=1}^{D}\sum_{t_{n}=1}^{D} \right)
  \prod_{m}
  T(K)_{x_{m} t_{m} x_{m-\hat{1}} t_{m-\hat{2}}},
\end{align}
where $D$ denotes the bond dimension of the four rank tensor $T\left( K \right)$
although the detailed definition of $T\left( K \right)$ is not discussed in this report.

Now the value of $Z\left( K \right)$ can be computed by taking the summation in eq.~\eqref{eq:Z_tensornetwork_approx}.
However, it is impossible to carry out the contraction of tensor indices in a large space-time extent.
In such a circumstance, it is useful to use a coarse-graining of the tensor network,
\textit{e.g.} the tensor renormalization group algorithm proposed by Levin and Nave~\cite{Levin:2006jai}.
In this study we use the simplest version of TRG though some variations have been developed so far.
A key ingredient of the TRG algorithm is using the singular value decomposition (SVD) to reduce the degrees of freedom.
In this report, we take the truncation order of the SVD to be equivalent to that of the initial tensor in eq.~\eqref{eq:Z_tensornetwork_approx}, the bond dimension $D$.

As shown in next section, our target quantity is the expectation value of the scalar field defined by
\begin{align}
  \label{eq:expectationvalue_phi}
  \left< \phi_{\tilde{n}} \right>
  = \frac{Z_{1}}{Z}
\end{align}
for a lattice site $\tilde{n}$.
Because of the translation invariance, $\left< \phi_{\tilde{n}} \right>$ takes the same value for any site $\tilde{n}$;
thus we simply write $\left< \phi \right>\left( =\left< \phi_{\tilde{n}} \right> \right)$ in the following.
The numerator in eq.\eqref{eq:expectationvalue_phi} is defined by
\begin{align}
  \label{eq:definition_Z1}
  Z_{1}
  = \left( \prod_{n} \int \mathrm{d} \phi_{n} \right) \phi_{\tilde{n}} e^{-S_{h}},
\end{align}
and its tensor network representation can be obtained in a similar way to the partition function although the detail is skipped here.

\section{Numerical results}
\label{sec:numericalresults}

In this section, we present our results of the dimensionless critical coupling.
We present numerical results of individual key steps: the thermodynamic limit, extracting the susceptibility,
a determination of the critical mass, and taking the continuum limit of the dimensionless critical coupling.
In the following we consider a system where the periodic boundary condition is imposed for all directions,
and we fix the number of discrete points for scalar fields $K$ as $256$.

\subsection{Thermodynamic limit and extraction of susceptibility}
\label{sec:thermodynamiclimit}

\begin{figure}[htbp]
  \begin{minipage}{0.5\hsize}
    \centering
    \includegraphics[width=\hsize]{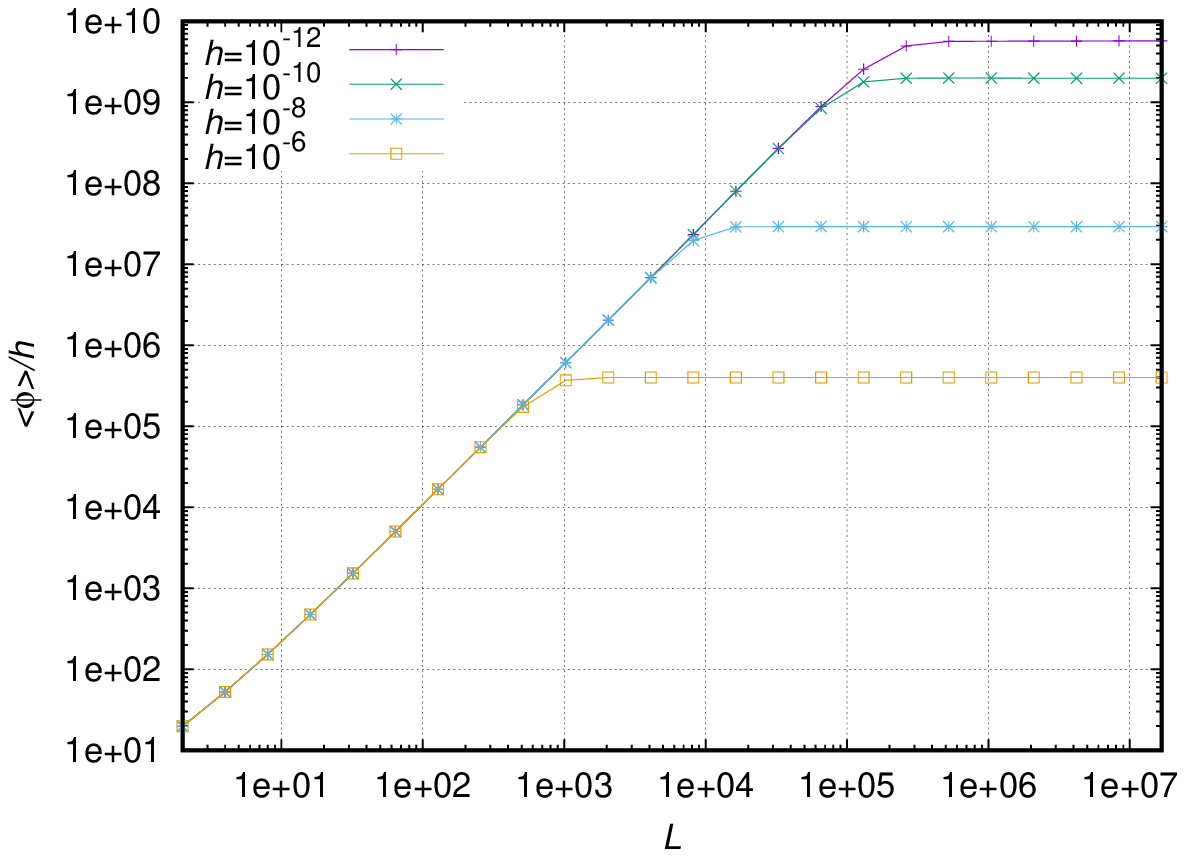}
    \caption{Thermodynamic limit of $\left< \phi \right>/h$ for
      $h\in\left[ 10^{-12}, 10^{-6} \right]$ at $\mu_{0}^{2} = -0.1006174$, $\lambda=0.05$, and $D=32$.
    }
    \label{fig:thermodynamiclimit}
  \end{minipage}
  \begin{minipage}{0.5\hsize}
    \centering
    \includegraphics[width=\hsize]{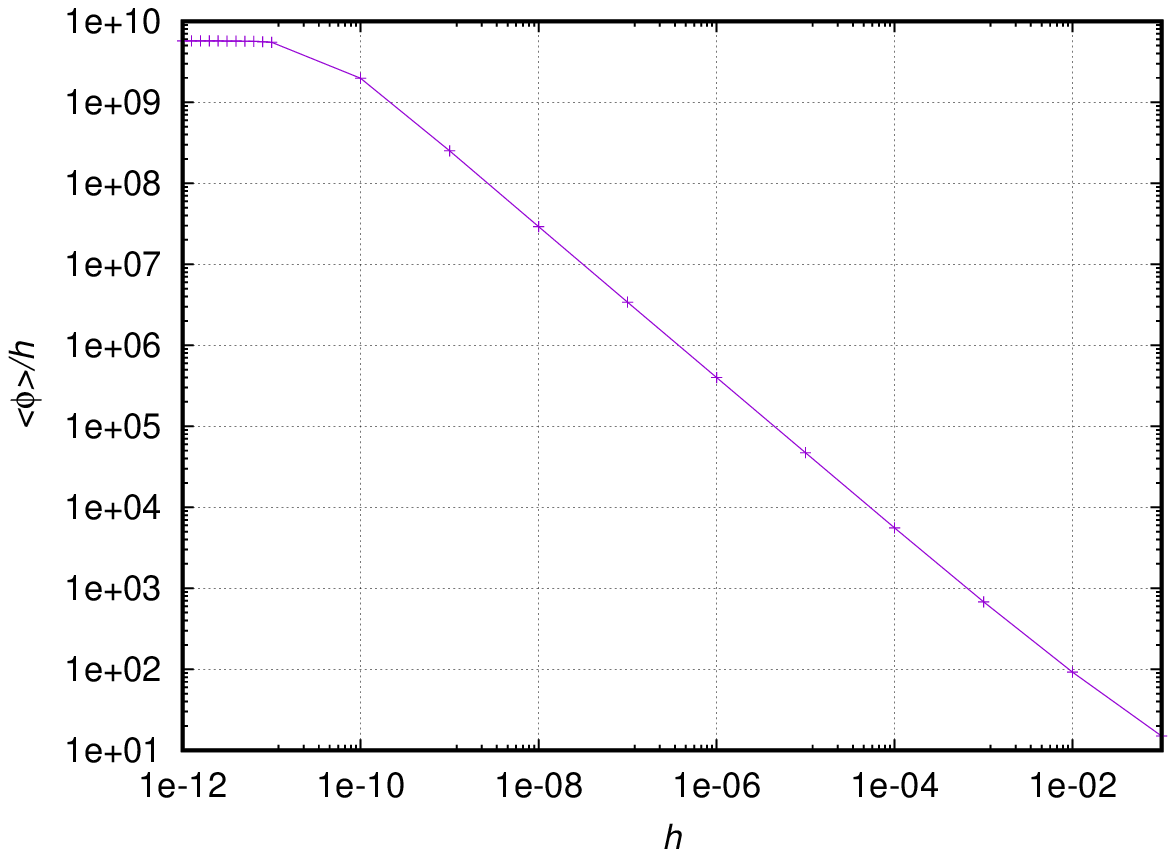}
    \caption{$\left< \phi \right>/h$ as a function of $h$ for $\mu_{0}^{2} = -0.1006174$, $\lambda=0.05$, and $D=32$ in the thermodynamic limit.
      The ratio at $h \to 0$ gives the susceptibility.}
    \label{fig:zerohlimit}
  \end{minipage}
\end{figure}

In fig.~\ref{fig:thermodynamiclimit} we
show $\left< \phi \right>/h$ as a function of $L$ for several values of $h$
with $\mu_{0}^{2} = -0.1006174$, $\lambda = 0.05$, $D=32$.
From the figure, one sees that the ratio becomes a constant in the extremely large volume region $L \ge 10^6$;
thus such region can be effectively considered as in the thermodynamic limit.

Figure~\ref{fig:zerohlimit}
shows $h$-dependence of $\left< \phi \right>/h$ in the thermodynamic limit
for the same parameter set ($\mu_{0}^{2}$, $\lambda$, and $D$) given above.
One can see that for sufficiently small $h \le 10^{-11}$ the ratio behaves as a constant;
thus the susceptibility $\chi$ can be obtained by using the relation $\left< \phi \right> \approx \chi \cdot h$.
Actually the values of $\left< \phi \right>/h$ in the range $h \in \left[ 10^{-12}, 10^{-11} \right]$ have a small fluctuation
that is hard to see at the scale of the figure.
Thus the obtained value of $\chi$ has an error estimated by using the amplitude of the fluctuation~\footnote{
  The error is very small compared to a fluctuation originated from $D$-dependence and almost does not affect on the following results.
}.
Although here we show the plots only for a particular choice of parameters ($\mu_{0}^{2}=-0.1006174$, $\lambda=0.05$, and $D=32$),
for the other parameter sets, we set up a proper range of the volume and the external field as well.
Our data of the susceptibility presented in the following are in the thermodynamic limit at the zero external field.

\subsection{Critical coupling}
\label{sec:criticalmass}

\begin{figure}[htbp]
  \begin{minipage}{0.5\hsize}
    \centering
    \includegraphics[width=\hsize]{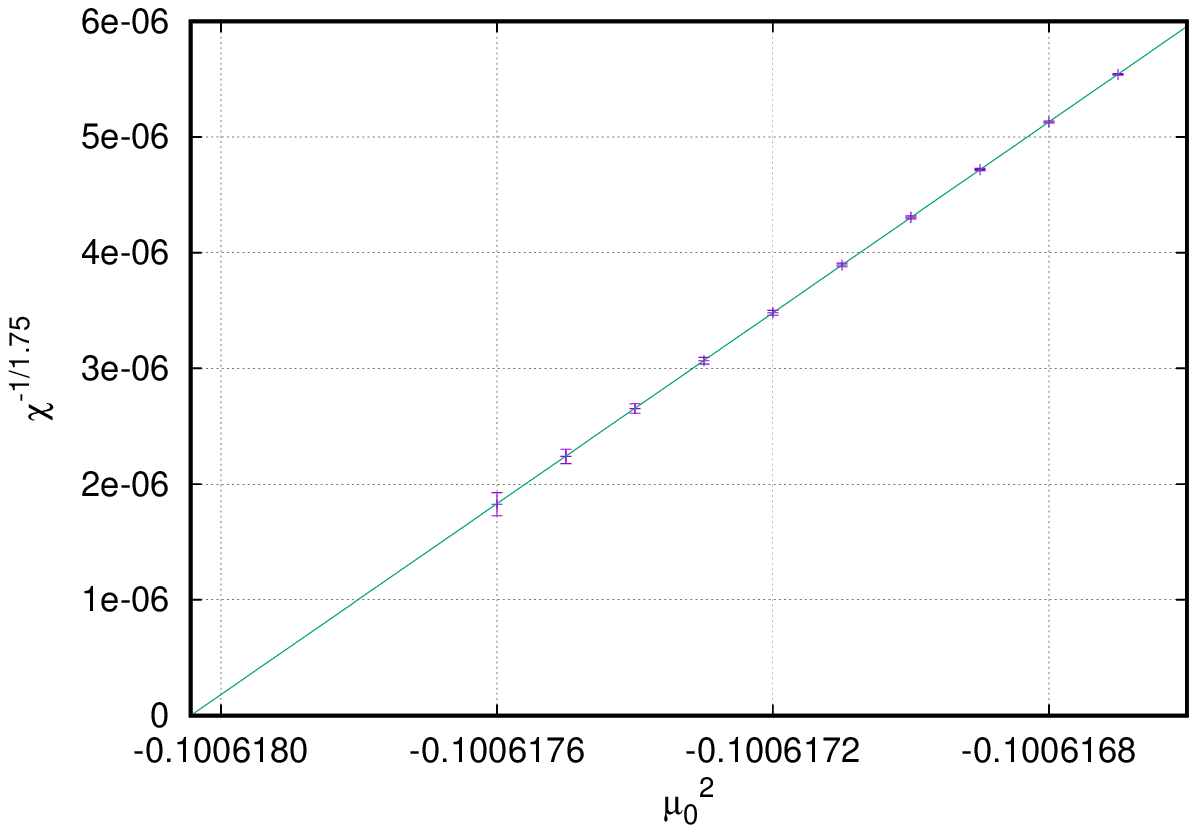}
    \caption{The deformed susceptibility as a function of $\mu_{0}^{2}$ at $\lambda=0.05$ and $D=32$.
      $\mu_{0, \mathrm{c}}^{2} = -0.1006180444(70)$ is obtained by a linear extrapolation with $\chi^{2}/\text{d.o.f.} \approx 0.0072$.
    }
    \label{fig:chi_linearfit}
  \end{minipage}
  \begin{minipage}{0.5\hsize}
    \vspace{-4ex}
    \centering
    \includegraphics[width=0.94\hsize,bb=0 0 504 360]{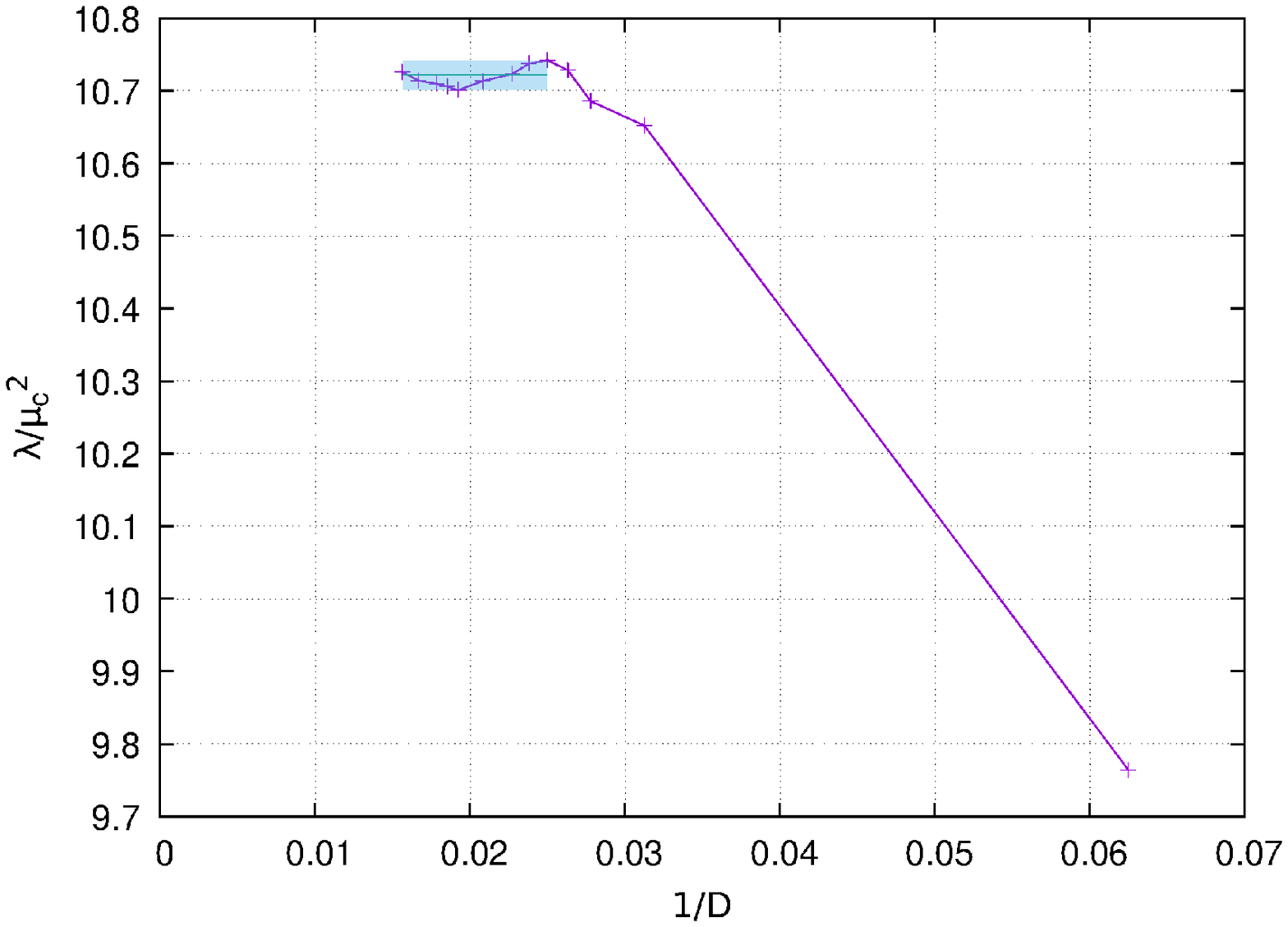}
    \caption{
      $D$-dependence of $\lambda/\mu_{\mathrm{c}}^{2}$
      at $\lambda = 0.05$.
      The error estimated from the fluctuation due to finite $D$ is also shown as the colored band.
    }
    \label{fig:criticalcoupling_D16-64}
  \end{minipage}
\end{figure}

Figure~\ref{fig:chi_linearfit} shows the deformed susceptibility
as a function of $\mu_{0}^{2}$ at $\lambda=0.05$ and $D=32$.
The critical mass is determined from the zero point of the fitting function
\begin{align}
  \label{eq:deformedfittingform}
  \chi^{-1/1.75}
  \propto \left| \mu_{0, \mathrm{c}}^{2} - \mu_{0}^{2} \right|^{\gamma/1.75}
\end{align}
with the fitting parameter, the critical (bare) mass square $\mu_{0, \mathrm{c}}^{2}$.
The $\phi^{4}$ theory is believed to belong to the two dimensional Ising universality class,
and the exact value of the critical exponent is known as $\gamma_{\mathrm{Ising}} = 1.75$;
then we here fix the exponent in eq.~\eqref{eq:deformedfittingform} as $\gamma = \gamma_{\mathrm{Ising}}$ to fit the data.
For $\lambda = 0.05$ and $D = 32$ we obtain $\mu_{0, \mathrm{c}}^{2} = -0.1006180444(70)$ with $\chi^{2}/\text{d.o.f.} \approx 0.0072$,
and this shows that fixing $\gamma = \gamma_{\mathrm{Ising}}$ is reasonable.
By repeating the same procedure for $0.005 \le \lambda \le 0.1$ and $16 \le D \le 64$,
we obtain critical masses, which is used in following analyses.

Figure~\ref{fig:criticalcoupling_D16-64} shows the bond dimension dependence of the dimensionless critical coupling $\lambda/\mu_{\mathrm{c}}^{2}$ at $\lambda=0.05$,
where $\mu_{\mathrm{c}}^{2}$ denotes the renormalized mass square~\footnote{
  Since the theory contains a single divergent diagram: the one-loop self energy,
  one has to perform the renormalization for the mass parameter.
  The coupling constant $\lambda$ is free of the renormalization since there is no corresponding divergent diagram in the two dimensional scalar theory.
}.
In the large $D$ region, $\lambda/\mu_{\mathrm{c}}^{2}$ exhibits an oscillating behavior.
We use the fluctuation to estimate the systematic error of $\lambda/\mu_{\mathrm{c}}^{2}$ as shown in fig.~\ref{fig:criticalcoupling_D16-64};
half of the difference between the maximum and the minimum value of $\lambda/\mu_{\mathrm{c}}^{2}$ in the shared area is adopted as the error.
As a central value, we simply quote an average between the maximum and minimum values.
We do the same procedure for the other values of $\lambda$.

\subsection{Continuum limit}
\label{sec:continuumlimit}

\begin{figure}[htbp]
  \centering
  \includegraphics[width=0.6\hsize]{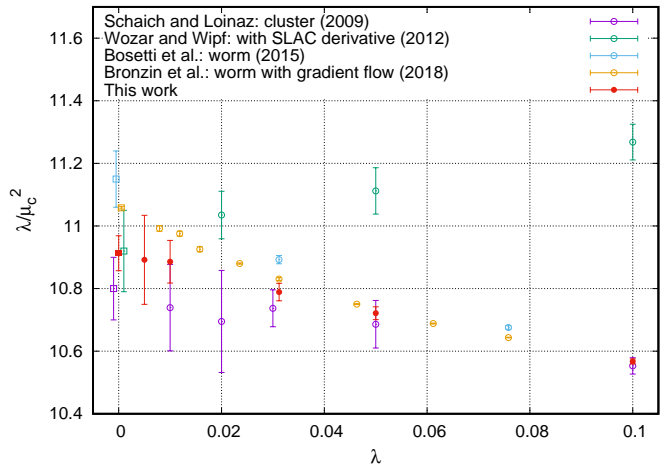}
  \caption{Comparison of the continuum extrapolation for $\lambda/\mu_{\mathrm{c}}^{2}$
    among the previous works (Schaich and Loinaz~\cite{Schaich:2009jk},
    Wozar and Wipf~\cite{Wozar:2011gu},
    Bosetti \textit{et al.}~\cite{Bosetti:2015lsa},
    and Bronzin \textit{et al.}~\cite{Bronzin:2018tqz}) and ours in the range $\lambda \le 0.1$.
    Square symbols at $\lambda = 0$ (horizontally shifted for the visibility) denote the continuum results of each method.
    Note that Wozar and Wipf used the SLAC derivative for scalar fields,
    so at non-zero $\lambda$ one cannot compare their result with others.
  }
  \label{fig:continuumlimit}
\end{figure}

Finally, by combining the results at all $\lambda$ values,
let us take the continuum limit ($\lambda \rightarrow 0$)
of the dimensionless critical coupling.
Figure~\ref{fig:continuumlimit} shows a comparison
among the recent Monte Carlo results and ours in the small $\lambda$ region.
To get the continuum value of the critical coupling $\left[ \lambda / \mu_{\mathrm{c}}^{2} \right]_{\mathrm{cont.}}$, we simply perform a linear extrapolation ($\chi^{2}/\text{d.o.f.} \approx 0.026$).
Our final result is given by
\begin{align}
  \label{eq:continuum_value}
  \left[ \frac{\lambda}{\mu_{\mathrm{c}}^{2}} \right]_{\mathrm{cont.}} = 10.913(56).
\end{align}
Although the smallest value of $\lambda=0.005$ is firstly reached by our current work,
the error bar is relatively larger compared with the latest Monte Carlo result around $\lambda \approx 0.01$.
Note that around $\lambda = 0.0312$ there are four data points of independent papers;
thus one can roughly compare their values.
As a result, the values of Schaich and Loinaz~\cite{Schaich:2009jk},
Bronzin \textit{et al.}~\cite{Bronzin:2018tqz},
and ours are roughly consistent with each other while
that of Bosetti \textit{et al.}~\cite{Bosetti:2015lsa} is relatively far away from the three results.
Table~\ref{tab:continuumresults} compiles the dimensionless critical coupling of the previous Monte Carlo works together with our work.

\captionsetup[figure]{width=\hsize, font=small}

\begin{table}[htbp]
  \centering
  \begin{tabular}{lcc}
    Method & Result & Year and Reference \\\hline\hline
    TRG with GH quadrature & $10.913(56)$ & 2018, this work \\
    MC worm with gradient flow & $11.058(4)$ & 2018,~\cite{Bronzin:2018tqz} \\
    MC worm & $11.15(6)(3)$ & 2015,~\cite{Bosetti:2015lsa} \\
    MC with SLAC derivative & $10.92(13)$ & 2012,~\cite{Wozar:2011gu} \\
    MC cluster & $10.8^{0.1}_{0.05}$ & 2009,~\cite{Schaich:2009jk}
  \end{tabular}
  \caption{
    The continuum results of the dimensionless critical coupling by previous Monte Carlo works and ours.
  }
  \label{tab:continuumresults}
\end{table}

\section{Summary and outlook}
\label{sec:Summary}

In this report, by using the TRG algorithm together with the new tensor network formulation,
we evaluate the dimensionless critical coupling of the two dimensional $\phi^{4}$ theory
and carry out its continuum extrapolation.
Our continuum value is roughly consistent with the previous results.
Although the result of the tensor network method does not have statistical errors,
our final result has relatively larger systematic error due to the finite bond dimension.
To reduce the error, one needs to increase the bond dimension,
but it requires more computational cost and memory usage.
Moreover, it is known that around the critical point,
the TRG algorithm suffers from the growth of the systematic errors.
In such a situation, however, as an alternative coarse-graining procedure,
the tensor network renormalization (TNR)~\cite{2015PhRvL.115r0405E} and loop-TNR~\cite{yang2017loop}
may be useful to obtain more precise results.
These methods, in principle, can be used in any model irrespective of the details of fields
as long as the system is defined on the square lattice.
Therefore in future it is interesting to apply these methods to the $\phi^{4}$ theory,
and we expect that the accuracy of the dimensionless critical coupling will be more improved.
We hope that such study will enhance the value of the tensor network method.

% \bibliographystyle{JHEP}
% \bibliography{/home/sakai/Workspace/references}
\providecommand{\href}[2]{#2}\begingroup\raggedright\endgroup

\end{document}